\documentclass[preprint,prd,showpacs,nofootinbib]{revtex4} 
 
\usepackage{amsmath}   

\begin{document}
\input{epsf}

\title{  Casimir Force between a Small Dielectric Sphere and a Dielectric Wall}  
  
\author{ V. Sopova}
 \email{svasilka@tufts.edu}  
\author{ L.H. Ford}
  \email{ford@cosmos.phy.tufts.edu}  
 
\affiliation{Institute of Cosmology \\ 
Department of Physics and Astronomy\\ 
Tufts University\\ 
Medford, Massachusetts 02155\\ }

\begin{abstract}
The possibility of repulsive Casimir forces between small metal spheres and
a dielectric half-space is discussed. We treat a model in which the spheres
have a dielectric function given by the Drude model, and the radius of the
sphere is small compared to the corresponding plasma wavelength. The
half-space is also described by the same model, but with a different plasma
frequency. We find that in the retarded limit, the force is
quasi-oscillatory. This leads to the prediction of stable equilibrium points
at which the sphere could levitate in the Earth's gravitational field. This
seems to lead to the possibility of an experimental test of the model.
The effects of finite temperature on the force are also studied, 
and found to be rather small at room temperature. However, thermally activated
transitions between equilibrium points could be significant at room temperature. 
\end{abstract}
 \maketitle 

\baselineskip=14pt

\section{Introduction }  

The interaction energy between a particle with non-dispersive polarizability 
$\alpha _{0}$ and a perfectly reflecting wall was found by Casimir and Polder
\cite{Polder} to be\footnote{Lorentz-Heaviside units with $c=\hbar =1$ will be 
used here.}

\begin{equation}
V_{CP}=-\frac{3\alpha _{0}}{32\pi ^{2}z^{4}},  \label{CP}
\end{equation}
where $z$ is the separation. One can assign a frequency spectrum 
$\hspace{ 0.01in}\sigma (\omega )$ to this potential and write \cite{Ford98} 

\begin{equation}
V_{CP}=\frac{\alpha _{0}}{16\pi ^{2}z^{3}}\int_{0}^{\infty }d\omega \hspace{
0.01in}\sigma (\omega ),  \label{Spectrum}
\end{equation}
where

\begin{equation}
\sigma (\omega )=(2\omega ^{2}z^{2}-1)\sin 2\omega z+2\omega z\cos 2\omega z.
\label{Sigma}
\end{equation}
The integrand, $\sigma (\omega )$, is an oscillatory function whose
amplitude \textit{increases }with increasing frequency. Nonetheless, the
integral can be performed using a convergence factor (\textit{e.g.}, insert
a factor of $e^{-\gamma \omega }$ and then let $\gamma \rightarrow 0$ after
integration). The result is the right hand side of Eq.~(\ref{CP}). It is
clear that massive cancellations have occurred, and that the area under an
oscillation peak can be much greater in magnitude than the final result.
This raises the possibility of tampering with this delicate cancellation,
and dramatically altering the magnitude and sign of the force.

In an earlier paper~\cite{Ford98}, henceforth I, 
this possibility was investigated for
small dielectric spheres near a perfectly reflecting wall. There it was
found that the force can be considerably larger than that given by 
Eq.~(\ref{CP}), and furthermore can be either attractive or repulsive. The force
was found to be quasi-oscillatory, with a period of the order of the plasma
wavelength of the material in the sphere. This effect can be understood as
a resonance phenomenon involving the vacuum modes and the plasma frequency
of the sphere. The calculations in I were restricted to the case of a
perfectly reflecting wall at zero temperature. 
 In the present
paper, we generalize these results to  the cases when
the wall has finite reflectivity and the temperature is non-zero.

In Sect.~\ref{sec:force}, a force on a small sphere near a dielectric wall is
computed at zero temperature. In  Sect.~\ref{sec:finiteT}, finite temperature
corrections are discussed. The results of the papers are summarized in 
Sect.~\ref{sec:sum}.

\section{Force on a Small Sphere near a Dielectric Wall}
\label{sec:force}

A small sphere of radius $a$ is placed a distance $z$ from a wall. Both the
sphere and the wall are composed of uniform material characterized with
dielectric function $\epsilon (\omega )$. We will take the dielectric
function to be that of the Drude model,

\begin{equation}
\epsilon (\omega )=1-\frac{\omega _{p}^{2}}{\omega (\omega +i\gamma )},
\label{drude}
\end{equation}
where $\omega _{p}$ is the plasma frequency and $\gamma $ is the damping
parameter. From now on, we take $\omega _{p}$ to be the plasma frequency of
the material in the sphere, and $\omega _{q}$ to be that of the interface.
We will use subscripts $w$ and $s$ to distinguish between other
parameters describing the wall and the sphere respectively. At present, we
assume that the temperature is zero.

The mean Casimir force on the sphere can be written as a Fourier transform:
\begin{equation}
\mathbf{F}=\frac{1}{2\pi }\int d\omega \,\mathbf{F}(\omega ),  \label{Fourier}
\end{equation}
where $\mathbf{F}(\omega )$ can be viewed as the contribution of a vacuum
mode of frequency $\omega $. This contribution can be found from essentially
classical considerations, as it is the normal component of the classical
force on the sphere when a plane wave of frequency $\omega $ is incident on
the sphere-interface system. In I, it was calculated in an
electric dipole approximation for the wave scattered by the sphere; the
force is found as an integral of the Maxwell stress tensor over the sphere.
In the present case, the extra complications come from the finite
reflectivity of the interface. However, these can be handled using the
dyadic Green's function techniques of Schwinger {\it et al}~\cite{Milton}.

The result for $\mathbf{F}$ (or $V$)
follows from 
\begin{equation}
\mathbf{F}(\omega )=-\frac{1}{2}\nabla \alpha _{r}(\omega )\left\langle
E^{2}\right\rangle _{\omega },  \label{Mode}
\end{equation}
where $\left\langle E^{2}\right\rangle $ is the (renormalized) expectation
value of the square of the electric field at the sphere's location and 
$\alpha _{r}(\omega )$ is the real part of the dynamic polarizability. This
expression is equivalent to the familiar result that the interaction energy
of an induced dipole with a static electric field $E_{0}$ is
\begin{equation}
V=-\frac{1}{2}\alpha _{0}E_{0}^{2},
\end{equation}
where $\alpha _{0}$ is the static polarizability of the particle. It turns
out that this expression can be applied to the dynamic case as well, e.g. by
expressing $\left\langle E^{2}\right\rangle _{\omega }$ in Eq.~(\ref{Mode}) as a
transverse spatial Fourier transform \cite{Vasilka}:
\begin{equation}
\left\langle E^{2}\right\rangle _{\omega }=
- {\rm Re}\left\{i\int (d\mathbf{k}_{\perp })\, 
\frac{1}{(2\pi )^{2}}\frac{1}{2\kappa }\left[ \omega ^{2}r+\left(
2k^{2}-\omega ^{2}\right) r^{\prime }\right] e^{-2\kappa z} \right\}\, .  
                           \label{Efield}
\end{equation}
where $\mathbf{k}_{\perp }$ is the transverse wavevector, and 
$k=\left| \mathbf{k}_{\perp }\right| $. 
The
reflection coefficients due to two polarization states of the electric field
vector, $r$ and $r^{\prime }$, are given by:
\begin{eqnarray}
r &\equiv &\frac{\kappa -\kappa _{1}}{\kappa +\kappa _{1}},  \label{Rcoeff}
\\
r^{\prime } &\equiv &\frac{\kappa \epsilon _{w}-\kappa _{1}}{\kappa \epsilon
_{w}+\kappa _{1}},  \label{R1coeff}
\end{eqnarray}
where $\kappa $ is defined by $\kappa ^{2}=k^{2}-\omega ^{2}$, and $\kappa
_{1}$ by $\kappa _{1}^{2}=k^{2}-\omega ^{2}\epsilon _{w}.$

A detailed discussion is given in the Appendix, but 
 the result
for the interaction potential ($\mathbf{F}=-\nabla V$) can be written as
\begin{equation}
V={\rm Re}\left\{\frac{i}{4\pi }\int_{-\infty }^{\infty }d\omega\, \alpha_{r}
 \int_{0}^{\infty }dkk\frac{1}{2\pi }\frac{1}{2\kappa }\left[ \omega
^{2}r+\left( 2k^{2}-\omega ^{2}\right) r^{\prime }\right] 
      e^{-2\kappa z} \right\}\,.                 \label{Vreal}
\end{equation}
This is similar to Eq.~(3.34) in Ref.~\cite{Milton}, which gives the interaction
energy between a molecule and a dielectric plate. In our case, the molecule
has been replaced by a dielectric sphere, and  the
polarizability of the sphere $\alpha $ has been replaced by its 
real part, $\alpha _{r}$. This is also a generalization of Eq.~(45) in I.
The treatment in I assumed no evanescent modes. However, the quantization of
the electromagnetic field in the presence of a dissipative dielectric requires
one to treat the frequency $\omega$ and wave number $k$ as independent
variables, effectively leading to evanescent wave contributions.
A detailed justification for the use of the real part of the dynamic
polarizability, $\alpha _{r}$, was given in I.

 The complex polarizability of the sphere is
given by
\begin{equation}
\alpha (\omega )=\alpha _{0}\frac{\epsilon _{s}(\omega )-1}{\epsilon
_{s}(\omega )+2},  \label{Alpha}
\end{equation}
where $\alpha _{0}$, the static polarizability, is given by $\alpha
_{0}=4\pi a^{3}$. If $\epsilon _{s}(\omega )$ is given by the Drude model,
Eq.~(\ref{drude}), then the real part of the polarizability is 
\begin{equation}
\alpha _{r}=\alpha _{0}\, \omega _{p}^{2}\frac{\omega _{p}^{2}-3\omega ^{2}}{  
(3\omega ^{2}-\omega _{p}^{2})^{2}+9\omega ^{2}\gamma _{s}^{2}}.
\label{AlphaR}
\end{equation}
This function has four poles in the complex $\omega$-plane, at 
$\pm \Omega \pm i \gamma _{s}/2$, where
\begin{equation}
\Omega = \frac{1}{6}\sqrt{12\omega _{p}^{2}-9\gamma _{s}^{2}} \, .
\end{equation}

\begin{figure}  
\begin{center}  
\leavevmode\epsfysize=6cm\epsffile{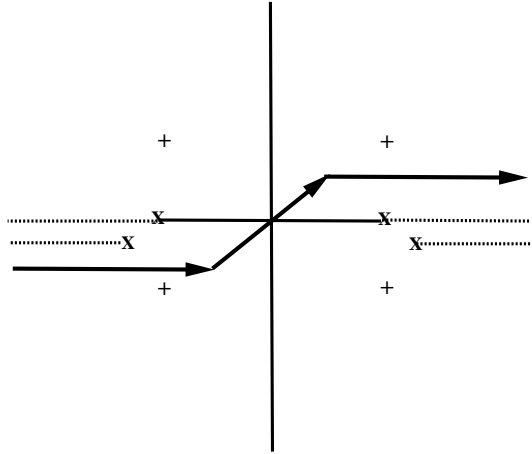}    
\end{center}  
\caption{The contour of integration for Eq.~(\ref{Vreal}) in the complex 
$\omega$-plane for fixed $k$ is illustrated. Here $X$ marks a branch point, and
a dotted line is its associated branch cut. The poles of the function
$\alpha _{r}(\omega)$ are marked with the $+$ symbol. 
When the contour is rotated to the imaginary axis, the poles in the first 
and third quadrants are the only singularities encountered.} 
\label{fig:contour}
\end{figure}

It will be convenient to deform the contour of integration in the 
$\omega$-plane, and isolate the residues of these poles. However, we must 
first consider the location of other possible singularities of the integrand
of Eq.~(\ref{Vreal}). There are branch points at the values of $\omega$
for which $\kappa = 0$ and $\kappa_1 = 0$. The former occur at 
$\omega = \pm k$. The latter are in the lower half $\omega$-plane. In the
limit that $\gamma_w \ll \omega_q$, they are located at approximately
\begin{equation}
\omega = \pm \sqrt{ \omega_q^2 +k^2} - 
i\, \frac{\gamma_w\, \omega_q^2}{2\,(\omega_q^2 +k^2)} \,.
\end{equation}
The dielectric function $\epsilon_w(\omega)$ has a pole at 
$\omega = - i \gamma_w$, but both of the reflection coefficients, $r$ and $r'$,
are regular at this point. Finally, there is a possibility of poles in the 
reflection coefficients at points at which $\kappa = -\kappa_1$ or
$\epsilon \,\kappa = -\kappa_1$. However, it may be shown that no such points
exist. The electric field Green's function should be defined by an integration
contour which goes beneath the singularities for ${\rm Re}\, \omega < 0$ and
above them for ${\rm Re}\, \omega > 0$, that is, Feynman boundary conditions. 
However, this does not include the
poles of $\alpha _{r}(\omega)$, which is not part of the electric field 
Green's function. Thus the contour of integration is as illustrated in
Fig.~\ref{fig:contour}.

We may now rotate the contour of integration to the imaginary $\omega$-axis,
so that $V$ can be written as the sum of an
integral over imaginary frequencies, $V_{j}$, and a contribution, $V_{p}$, 
from the residues of the pole in $\alpha _{r}$:

\begin{equation}
V=V_{j}+V_{p},  \label{Vsum}
\end{equation}
with

\begin{equation}
V_{j}=-\frac{1}{8\pi ^{2}}\int_{0}^{\infty }d\zeta \alpha _{r}(i\zeta
)\int_{0}^{\infty }dkk\frac{1}{\kappa }\left[ -\zeta ^{2}\overline{r}+\left(
2k^{2}+\zeta ^{2}\right) \overline{r^{\prime }}\right] e^{-2\kappa z},
\label{Vj}
\end{equation}
and
\begin{equation}
V_{p}=\frac{\alpha _{0}\,\omega_{p}^2}{48\pi \Omega}\; {\rm Re}
\int_{0}^{\infty }dkk\frac{1}{\kappa (\omega _{1})}\left[ \omega _{1}^{2}\, 
\overline{r}(\omega _{1})+\left( 2k^{2}-\omega _{1}^{2}\right) \overline{
r^{\prime }}(\omega _{1})\right] e^{-2\kappa (\omega _{1})z},  \label{Vp}
\end{equation}
where $\omega _{1}$ is the pole of $\alpha _{r\text{ }}$in the first quadrant,

\begin{equation}
\omega _{1}=\frac{1}{6}\sqrt{12\omega _{p}^{2}-9\gamma _{s}^{2}}+\frac{1}{2}
i\gamma _{s},  \label{Pole}
\end{equation}
and $\overline{r}$ is defined as

\begin{equation}
\overline{r}=\frac{r(\gamma _{w})+r(-\gamma _{w})}{2},  \label{Ave}
\end{equation}
and similarly for $\overline{r^{\prime }}$. The first term in Eq.~(\ref{Ave})
arises from the pole in the first quadrant, and the second term from that in
the third quadrant, and we have used the fact that the Drude model dielectric
function satisfies $\epsilon(-\omega,\gamma) = \epsilon(\omega,-\gamma)$.

By introducing polar coordinates $u$ and $\theta $ ( $\zeta =u\cos \theta ,$ 
$k=u\sin \theta $ ), and subsequently taking $t\equiv \cos (\theta )$, 
Eq.~(\ref{Vj}) becomes

\begin{equation}
V_{j}=-\alpha _{0}\frac{\omega _{p}^{2}}{8\pi ^{2}}\int_{0}^{\infty
}du\int_{0}^{1}dt\frac{3u^{2}t^{2}+\omega _{p}^{2}}{(3u^{2}t^{2}+\omega
_{p}^{2})^{2}-9u^{2}t^{2}\gamma _{s}^{2}}u^{3} \;{\rm Re} \left[ -t^{2}\overline{r} 
+\left( 2-t^{2}\right) \overline{r^{\prime }}\right] e^{-2uz}.  \label{Vjut}
\end{equation}
The numerical evaluation of the integral is done in the limit $\gamma
_{s}\rightarrow 0$, $\gamma _{w}\rightarrow 0$ (plasma model regime). In
this case we can write the coefficients $r$ and $r^{\prime }$ (in place of 
$\overline{r}$ and $\overline{r^{\prime }}$) , as defined in Eq.~(\ref{R1coeff}),
in terms of the new variables as

\begin{mathletters}    
\begin{eqnarray}
r &=&\frac{u-\sqrt{u^{2}+\omega _{q}^{2}}}{u+\sqrt{u^{2}+\omega _{q}^{2}}},
\label{Rpolar} \\
r^{\prime } &=&\frac{u^{2}t^{2}+\omega _{q}^{2}-ut^{2}\sqrt{u^{2}+\omega
_{q}^{2}}}{u^{2}t^{2}+\omega _{q}^{2}+ut^{2}\sqrt{u^{2}+\omega _{q}^{2}}}.
\label{R1polar}
\end{eqnarray}
\end{mathletters}  

\begin{figure}  
\begin{center}  
\leavevmode\epsfysize=6cm\epsffile{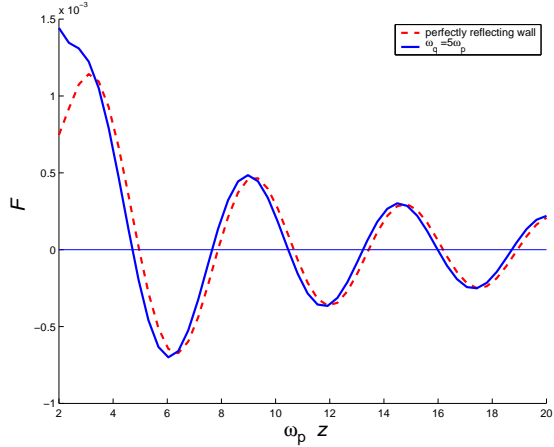}   
\end{center}  
\caption{The force, in units of $\protect 
\omega_{p}^{5}\,\alpha_0$, is given for the case $\protect\omega _{q}=5\protect 
\omega _{p}$. Repulsion corresponds to $F>0$.
 Here the result is quite close to that for the perfectly
reflecting wall (dashed line). } 
\label{fig:T0_r5} 
\end{figure}

\begin{figure}  
\begin{center}  
\leavevmode\epsfysize=6cm\epsffile{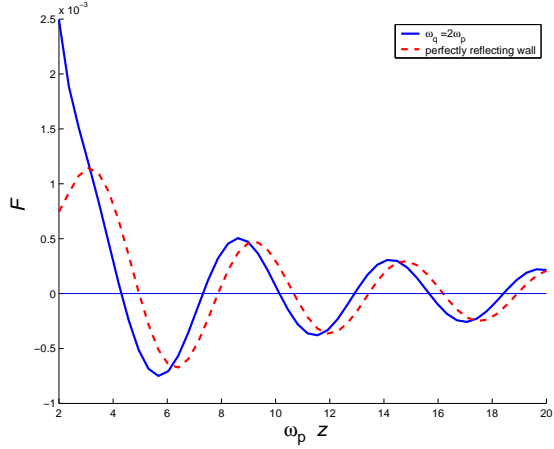}    
\end{center}  
 \caption{The force is given for
the case $\protect\omega _{q}=2\protect\omega _{p}$. Here the finite
reflectivity has caused a shift in the locations of the force maxima and
minima, but little change of the magnitude of the maximum force. }
\label{fig:T0_r2}  
\end{figure}

\begin{figure}  
\begin{center}  
\leavevmode\epsfysize=6cm\epsffile{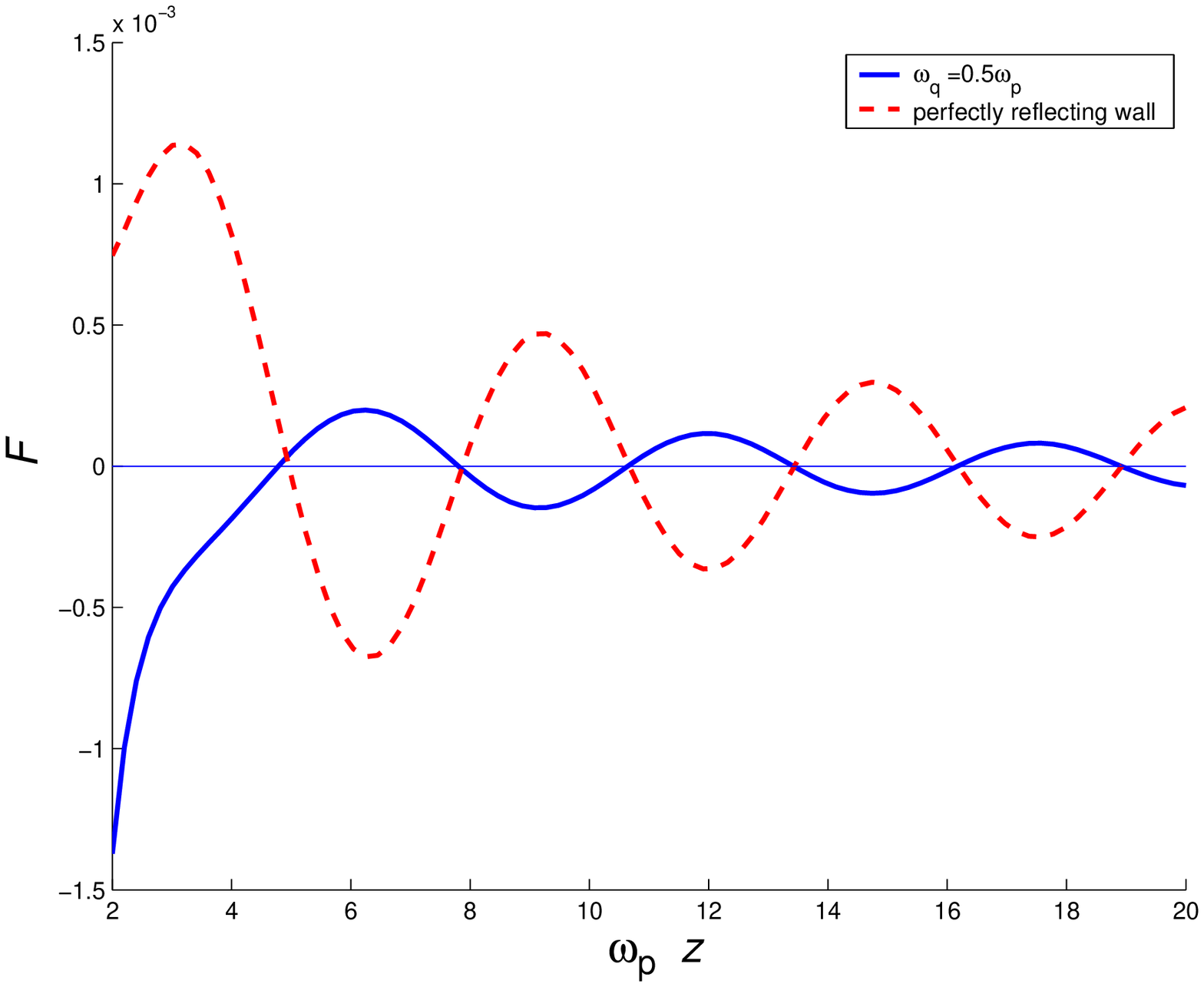}   
\end{center}  
 \caption{The force is given for
the case $\protect\omega _{q}=0.5\protect\omega _{p}$. Besides a shift in
the locations of the force maxima and minima, the finite reflectivity has
caused a reduction in the magnitude of the maximum force. }
\label{fig:T0_r05} 
\end{figure}

In Figures~\ref{fig:T0_r5} - \ref{fig:T0_r05}, 
the force on the sphere ($\mathbf{F}=-\nabla \left(
V_{j}+V_{p}\right) $) in units $\omega_{p}^{5}\,\alpha_0$ is given for three
different ratios $\omega _{q}/\omega _{p}$. The $V_{p}$ term is typically
dominant, and gives rise to a quasi-oscillatory contribution to the force,
as seen in the figures. This quasi-oscillatory behavior is in
contrast to the monotonic attractive force in the case of an atom in its
ground state interacting with an interface \cite{Polder}, \cite{Milton}.
However, it is similar to the force between an atom in an excited state and
an interface \cite{Barton}, which is also quasi-oscillatory. In the present
case, the oscillations arise from a resonance at $\omega = \omega_1 \approx
\sqrt{3}\omega_{p}/3$, if $\gamma_s \ll \omega_p$,
 as seen from Eq.~(\ref{Pole}). The appearance of the real part of the
polarizability is crucial for this feature of the result. The force is
alternatively attractive and repulsive as a function of separation, and this
leads to a sequence of stable equilibrium positions for the sphere. These
are the zeros of the function $\mathbf{F}(z)$ with negative slope. This
seems to lead to the possibility of an experimental test of the model in
which the sphere could levitate in the Earth's gravitational field at
positions where $\mathbf{F}$ cancels the sphere's weight, as discussed in I.
Note that $F$ is defined so that $F>0$ is repulsion, and $F<0$ is attraction.
Also note that because the present paper adopts Lorentz-heaviside units, in 
contrast to the Gaussian units used in I, the scale in the figures differs 
by a factor of $4\pi$ from Fig. 7 in I. 

Of particular interest is the effect of the finite reflectivity of the
interface, which is described by its plasma frequency, $\omega _{q}$. As
expected, when $\omega _{q}\gg \omega _{p}$, the results agree with those
given in I for a perfectly reflecting wall. As the ratio $\omega
_{q}/\omega _{p}$ decreases, the force oscillations shift their phase and
eventually decrease in amplitude, as seen in the figures. 
In Figure~\ref{fig:T0_r5} we
see that if $\omega _{q}$ is larger than a few times $\omega _{p}$, the
perfectly conducting result is a very good approximation (dashed line). Even 
when $\omega _{q}<\omega _{p}$, as in Figure~\ref{fig:T0_r05}, 
one still finds the
quasi-oscillatory behavior of the force. The amplitude of the force
decreases, as expected, since the effect should disappear as $\omega
_{q}\rightarrow 0$.

In the small distance limit, $a\ll z\ll \omega _{p}^{-1}$, the expression
for the total force on the particle in the plasma model regime becomes

\begin{equation}
\mathbf{F}=-\frac{d}{dz}(V_{j}+V_{p})\approx -\frac{\alpha _{0}}{4\pi }\frac{ 
3\sqrt{2}}{8}\frac{\omega _{p}^{2}\omega _{q}}{2\omega _{p}^{2}-3\omega
_{q}^{2}}\frac{1}{z^{4}}.  \label{Fsmallz}
\end{equation}
We see that the force diverges near the interface, and moreover its sign
depends on the ratio $\omega _{p}/\omega _{q}$, which seems to be an
artifact of the assumption of a perfectly smooth interface. As discussed
in Ref.~\cite{Vasilka}, dispersion alone is not sufficient to render
mean squared electromagnetic fields finite in the limit that one approaches 
such an interface.

\subsection{Perfectly reflecting Wall}

In the limit $\epsilon _{w}\rightarrow \infty $, as seen from (\ref{Rcoeff})
and (\ref{R1coeff}), $r\rightarrow -1$ , and $r^{\prime }\rightarrow 1$ . In
this case (\ref{Vjut}) can be analytically integrated over $t$ yielding:

\begin{eqnarray}
V_{j}= & &\alpha _{0}\frac{\omega _{p}^{2}}{4\pi ^{2}}\frac{1}{\sqrt{3}\sqrt{
4\omega _{p}^{2}-3\gamma _{s}^{2}}}\int_{0}^{\infty }du\, u^{2} \nonumber \\
  &\times& \left[ \arctan
\left( \frac{\sqrt{3}\left( -2u+\gamma _{s}\right) }{\sqrt{4\omega
_{p}^{2}-3\gamma _{s}^{2}}}\right)   
  -\arctan \left( \frac{\sqrt{3}\left(
2u+\gamma _{s}\right) }{\sqrt{4\omega _{p}^{2}-3\gamma _{s}^{2}}}\right) 
\right] e^{-2uz},  \label{VjPerf}
\end{eqnarray}
and (\ref{Vp}), after analytical evaluation, becomes:

\begin{equation}
V_{p}=-\frac{\alpha _{0}}{4\pi }\frac{\omega _{p}^{2}}{24 \Omega }\frac{1}{ 
z^{3}}\left( 2\omega _{1}^{2}z^{2}+2i\omega _{1}z-1\right) e^{2i\omega_{1} z}.
\label{VpPerf}
\end{equation}

It can be shown that these results agree with the ones in I,
namely we find that the total force on the sphere in this limit becomes:

\begin{equation}
F=J+P,  \label{Fjp}
\end{equation}
where

\begin{equation}
J=-\frac{d}{dz}V_{j}=-\alpha _{0}\frac{\omega _{p}^{2}}{16 \pi^2 z^{4}} 
\int_{0}^{\infty }d\xi \frac{\left( 3\xi ^{2}+\omega _{p}^{2}\right) \left(
4z^{3}\xi ^{3}+6z^{2}\xi ^{2}+6z\xi +3\right) }{\left( 3\xi ^{2}+\omega
_{p}^{2}\right) ^{2}-9\xi ^{2}\gamma _{s}^{2}}e^{-2z\xi },  \label{J}
\end{equation}
and

\begin{eqnarray}
P&=&{\rm Re}\left( -\frac{d}{dz}V_{p}\right) \nonumber \\
 &=& -\alpha _{0}\frac{\omega
_{p}^{2}}{192 \pi\Omega }\frac{1}{z^{4}}e^{-\gamma _{s}z}\left[ 2\Omega z\left(
4\Omega ^{2}z^{2}-3\gamma _{s}^{2}z^{2}-6\gamma _{s}z-6\right) \sin 2\Omega
z+\right.  \nonumber \\
& & \left. \left( 12\gamma _{s}\Omega ^{2}z^{3}-\gamma _{s}^{3}z^{3}+12\Omega
^{2}z^{2}-3\gamma _{s}^{2}z^{2}-6\gamma _{s}z-6\right) \cos 2\Omega z\right]\, .
  \label{P}
\end{eqnarray} 
As seen here, the plasma model is a good approximation as long as 
$\gamma _{s}z\ll 1$. Otherwise, Eq.~(\ref{P}) can yield significant corrections
for the more distant equilibrium positions, but not for the first
several peaks. We expect this to be true for the case of finite conductivity
of the wall as well, as long as $\gamma _{s} \ll \omega _{p}$, and $\gamma_{w}
\ll \omega _{q}$.

\section{Finite Temperature Corrections}
\label{sec:finiteT}

I\smallskip \smallskip n the case of nonzero temperature, Eqs.~(\ref{Vj}) 
and (\ref{Vp}) have to be modified. We write Eq.~(\ref{Vj}) as a
Fourier series instead of Fourier transform by the substitution \cite{Milton}

\begin{equation}
\zeta ^{2}\rightarrow \zeta _{n}^{2}=\frac{4\pi ^{2}n^{2}}{\beta ^{2}} 
,\qquad \int_{0}^{\infty }\frac{d\zeta }{2\pi }\longrightarrow \frac{1}{
\beta }\sum_{n=0}^{\infty }{}^{\prime }.  \label{Zeta}
\end{equation}
The prime is a reminder to count the $n=0$ term with half weight, and $\beta
=1/kT$. Thus, in the limit $\gamma _{s},\gamma _{w}\approx 0$:

\begin{equation}
V_{j}= -\frac{\alpha _{0}\omega _{p}^{2}}{4\pi \beta } \;{\rm Re} 
\sum_{n=0}^{\infty }{}^{\prime }\frac{1}{3\zeta _{n}^{2}+\omega _{p}^{2}}
\int_{0}^{\infty } \frac{dk \,k}{\kappa}
\left[ -\zeta _{n}^{2}r+\left( 2k^{2}+\zeta
_{n}^{2}\right) r^{\prime }\right] e^{-2\kappa z},  \label{VjT}
\end{equation}
where $r$ and $r^{\prime }$, using Eqs.~(\ref{Rcoeff}), (\ref{R1coeff}) 
 and (\ref{drude}) can be written as:

\begin{mathletters}  
\label{RcoeffsT}  
\begin{eqnarray}
r &\equiv &\frac{\kappa -\sqrt{\kappa ^{2}+\omega _{q}^{2}}}{\kappa +\sqrt{
\kappa ^{2}+\omega _{q}^{2}}},  \label{RcoefT} \\
r^{\prime } &\equiv &\frac{\kappa \left( \zeta _{n}^{2}+\omega
_{q}^{2}\right) -\zeta _{n}^{2}\sqrt{\kappa ^{2}+\omega _{q}^{2}}}{\kappa
\left( \zeta _{n}^{2}+\omega _{q}^{2}\right) +\zeta _{n}^{2}\sqrt{\kappa
^{2}+\omega _{q}^{2}}}.  \label{RcoeffT1}
\end{eqnarray}
We modify Eq.~(\ref{Vp}) by inserting a factor $\left( 1+\frac{2}{e^{\beta
\omega }-1}\right) $ to account for the thermal energy. This factor reflects
the fact that at zero temperature, each mode has an energy of $\frac{1}{2}
\omega $; at finite temperature, there is an additional thermal energy of $
1/\left( e^{\beta \omega }-1\right) $. The result is

\end{mathletters}  
\begin{equation}
V_{p}=\frac{\alpha _{0}\,\omega_{p}^{2}}{24\pi \Omega }
\;{\rm Re} \left\{\left( \frac{1
}{e^{\beta \omega_1 }-1}+\frac{1}{2}\right) \int_{0}^{\infty }dkk\frac{1}{
\kappa (\omega _{1})}\left[ \omega _{1}^{2}r(\omega _{1})+\left(
2k^{2}-\omega _{1}^{2}\right) r^{\prime }(\omega _{1})\right] e^{-2\kappa
(\omega _{1})z}\right\}\, .  \label{VpT}
\end{equation}

\begin{figure}  
\begin{center}  
\leavevmode\epsfysize=6cm\epsffile{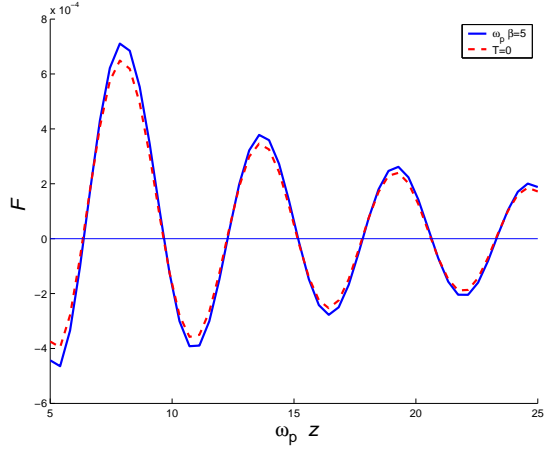}   
\end{center}  
\caption{The force is given for the case
$\omega_q = \omega_p$ and $\omega_p \beta =5$. The effect of the finite 
temperature is a slight increase in the magnitude of the force.} 
\label{fig:Force_b5}   
\end{figure}

The electric dipole approximation requires that the radius of the sphere be
small compared to the dominant wavelengths involved, so our results require
that $a\ll \beta $. So long as the temperature stays within this range of
validity, the thermal correction enhances the magnitude of the force, with
no shift in the positions of the force maxima and minima as illustrated in 
Figure~\ref{fig:Force_b5} for case $\omega _{p}\beta =5$. 
In most of the cases of interest, 
$\omega _{p}$ is in the ultraviolet range, so $\omega _{p}\beta \gg 1$ at
room temperature. The temperature is also limited by the melting point of
the material of the wall.

Although the finite-temperature corrections to the force are small, there
is another thermal effect which can be more significant. This is the
possibility of thermally activated transitions between the stable equilibrium
points, which become important when the required energy is of order $k T$.
We can estimate this effect by noting that for the case $\omega_q =2 \omega_p$,
this energy is approximately
\begin{equation}
W = \int_{4/\omega_p}^{7/\omega_p} f \, dz \approx 
               1.4 \times 10^{-3}\, \alpha_0 \,\omega_p^4 \,,
\end{equation}
which can be expressed as
\begin{equation}
W = 2200K \left(\frac{a}{20 nm}\right)^3 \left(\frac{\omega_p}{10 eV}\right)^4
 = 280K \left(\frac{a}{10 nm}\right)^3 \left(\frac{\omega_p}{10 eV}\right)^4 \,.
\end{equation}
Note that $10 eV = 1/20 nm$, so for $\omega_p \approx 10 eV$ the upper limit
on the size of a sphere that would fit inside of a particular potential minimum
is about $20 nm$.
In this case, the thermal activation effect would be small at room temperature.
However, for a smaller sphere, e.g., $a = 10 nm$, thermal activation would
be noticeable at room temperature.

\section{Summary}
\label{sec:sum}

We have investigated a model for the amplification of vacuum fluctuations,
in which a resonance near the plasma frequency significantly increases the
magnitude of the force compared to the non-dispersive case. In particular,
we have examined the effects of finite reflectivity of the wall and of
finite temperature. Finite reflectivity can decrease the magnitude of the
effect, but so long as $\omega _{q}$ is of the order of, or larger than $ 
\omega _{p}$, the perfectly reflecting results are qualitatively correct.
Finite temperature can enhance the magnitude of the effect, but with typical
plasma frequencies in the ultraviolet range, the correction to the force
 at room temperature is extremely small. However, thermal activation
can be significant, especially for smaller spheres.

The key prediction of this model is a force that is alternatively attractive
and repulsive as a function of separation.  The ratio of
the peak force to that of gravity on a sphere near an interface with
$\omega_q > \omega_p$ is approximately given by
 \begin{equation}
\frac{F_{max}}{F_g} \approx 27\, \left(\frac{\omega_p}{1 eV}\right)^4\,
\left(\frac{1 \mu m}{z}\right)\, \left(\frac{1 g/cm^3}{\rho}\right) \,
e^{-5\,(\gamma_s/ 1eV)\,(z/ 1 \mu m)} \,,
\end{equation}
where $\rho$ is the mass density of the sphere. 
This suggests a possibility of an
experimental test of the model in which the sphere could levitate in the
Earth's gravitational field at equilibrium positions. If forces whose
magnitudes are large compared to those give by Eq.~(\ref{CP}) with $\alpha
_{0}\approx 4 \pi a^{3}$ are found, the result can be interpreted as a form of
amplification of vacuum fluctuations. The spheres need to be small 
($a\omega_{p}\alt 1$), so experiments on metal sphere would have to involve 
spheres with radii in approximately the $10nm$ to $100nm$ range. In the case
of gold spheres, for example, ($\omega_p = 9 eV$), the sphere radius would have 
to be less than $20 nm$ and preferably less than about $10 nm$. 

One might also ask if a similar amplification is possible for the Casimir
force between perfectly reflecting parallel plates. The spectrum of the
Casimir force in this case was discussed by 
Hacyan, \textit{et al}~\cite{Hacyan}, and one of the present 
authors~\cite{Ford88}, and found to be
quasi-oscillatory, as in the case of the Casimir-Polder potential. However,
as was shown by Lifshitz \cite{Lifs}, the force
between a pair of dielectric half-spaces is always attractive and no larger
in magnitude than the Casimir force. Thus upsetting the cancellation seems
to be more difficult for half-spaces, and suggests that the small sphere
approximation may be crucial. to obtain a quasi-oscillatory force. For 
larger objects, there may be a cancellation of the contributions of different
spatial regions.

\begin{acknowledgments}
  This work was supported in part by the National
Science Foundation under Grant PHY-0244898.
\end{acknowledgments}

\appendix
\section{Derivation for the Force on a small Sphere near a Wall}

We can obtain Eq.~(\ref{Mode}) by using Eq.~(11) in I as a
starting point for the force on the sphere\footnote{This expression is obtained
 by integrating the Maxwell stress tensor over a
spherical surface just outside the sphere.}:

\begin{equation}
F^{i}=\frac{2}{3}p^{j}\partial _{j}E^{i}+\frac{1}{3}p_{j}\partial ^{i}E^{j}+ 
\frac{2}{3}\left( \dot{\mathbf{p}}\times \mathbf{B}\right) ^{i},
\label{FFord}
\end{equation}
where $\mathbf{p}$ is the dipole moment associated with the sphere, and 
$\mathbf{E}$ and $\mathbf{B}$ are the mean electric and magnetic field
vectors at the sphere's location. We take $\mathbf{p}$ to be linearly
related to $\mathbf{E}$: $\mathbf{p=}\alpha \mathbf{E}$. Using the Maxwell
equation, $\dot{\mathbf{E}}=\mathbf{\nabla }\times \mathbf{B}$, we get for the
last term in Eq.~(\ref{FFord}):

\begin{equation}
\frac{2}{3}\left( \dot{\mathbf{p}}\times \mathbf{B}\right) ^{i}=\frac{2}{3}
\alpha \left[ \left( \mathbf{\nabla }\times \mathbf{B}\right) \times \mathbf{
B}\right] ^{i}=\frac{2}{3}\alpha \left[ -\left( \partial ^{i}B^{k}\right)
B_{k}+\left( \partial ^{k}B^{i}\right) B_{k}\right] .  \label{F3term}
\end{equation}
Thus Eq.~(\ref{FFord}) becomes:

\begin{equation}
F^{i}=\frac{1}{3}\alpha \left[ 2E^{j}\partial _{j}E^{i}+E_{j}\partial
^{i}E^{j}+2B_{j}\partial ^{j}B^{i}-2B_{j}\partial ^{i}B^{j}\right]
\label{FFord1}
\end{equation}
Now, we replace the field products with their appropriate expectation values
for $i=z$. The expectation values of the electric and magnetic fields can be
expressed through the Green's dyadic as in \cite{Milton}:

\begin{equation}
\frac{i}{\hbar }\left\langle E_{j}(\mathbf{r})\,E_{k}(\mathbf{r}^{\prime
})\right\rangle =\Gamma _{jk}(\mathbf{r},\mathbf{r}^{\prime },\omega ),
\label{EfieldG}
\end{equation}
and
\begin{equation}
\frac{i}{\hbar }\left\langle B_{j}(\mathbf{r})\,B_{k}(\mathbf{r}^{\prime
})\right\rangle =\epsilon _{jlm}\epsilon _{knp}(\partial _{l}\partial
_{n}/\omega ^{2}\,)\Gamma _{mp}(\mathbf{r},\mathbf{r}^{\prime },\omega ).
\label{MfieldG}
\end{equation}
Quantities such as 
$\langle E_{j}(\mathbf{r})\,E_{k}(\mathbf{r}^{\prime})\rangle$ must be real, 
so we need to take a real part, which will only be done explicitly in the 
final expressions. 
Some components of $\mathbf{\,}\overleftrightarrow{\mathbf{\Gamma }}$ are
(here $\mathbf{k}_{\perp }$ is chosen to point along the $+x$ axis):

\begin{eqnarray}
\Gamma _{xx}\mathbf{(\mathbf{r},\mathbf{r}^{\prime },}\omega \mathbf{)} &=& 
\mathbf{\int }d\mathbf{\mathbf{k}_{\bot }\;}\frac{1}{(2\pi )^{2}}e^{i\mathbf{
k}_{\bot }(\mathbf{r}-\mathbf{r}^{\prime })_{\bot }} \nonumber \\
& \times& \left[ -\frac{1}{
\epsilon }\delta (z-z^{\prime })+\frac{1}{\epsilon }\frac{\partial }{
\partial z}\frac{1}{\epsilon ^{\prime }}\frac{\partial }{\partial z^{\prime }
}\left( \frac{e^{-\kappa \mid z-z^{\prime }\mid }+r^{\prime }e^{-\kappa
(z+z^{\prime })}}{2\kappa }\right) \right] ,  \label{Gxx} \\
\Gamma _{yy}\mathbf{(\mathbf{r},\mathbf{r}^{\prime },}\omega \mathbf{)} &=&
\mathbf{\int }d\mathbf{\mathbf{k}_{\bot }\;}\frac{1}{(2\pi )^{2}}e^{i\mathbf{
k}_{\bot }(\mathbf{r}-\mathbf{r}^{\prime })_{\bot }}\omega ^{2}\left( \frac{
e^{-\kappa \mid z-z^{\prime }\mid }+re^{-\kappa (z+z^{\prime })}}{2\kappa }
\right) ,  \label{Gyy} \\
\Gamma _{zz}\mathbf{(\mathbf{r},\mathbf{r}^{\prime },}\omega \mathbf{)} &=&
\mathbf{\int }d\mathbf{\mathbf{k}_{\bot }\;}\frac{1}{(2\pi )^{2}}e^{i\mathbf{
k}_{\bot }(\mathbf{r}-\mathbf{r}^{\prime })_{\bot }}  \nonumber \\
& \times& \left[ -\frac{1}{%
\epsilon }\delta (z-z^{\prime })+\frac{k^{2}}{\epsilon \epsilon ^{\prime }}
\left( \frac{e^{-\kappa \mid z-z^{\prime }\mid }+r^{\prime }e^{-\kappa
(z+z^{\prime })}}{2\kappa }\right) \right] ,  \label{Gzz} \\
\Gamma _{xz}\mathbf{(\mathbf{r},\mathbf{r}^{\prime },}\omega \mathbf{)} &=&
\mathbf{\int }d\mathbf{\mathbf{k}_{\bot }\;}\frac{1}{(2\pi )^{2}}e^{i\mathbf{
k}_{\bot }(\mathbf{r}-\mathbf{r}^{\prime })_{\bot }}i\frac{k}{\epsilon
\epsilon ^{\prime }}\frac{\partial }{\partial z}\left( \frac{e^{-\kappa \mid
z-z^{\prime }\mid }+r^{\prime }e^{-\kappa (z+z^{\prime })}}{2\kappa }\right)
,  \label{Gxz} \\
\Gamma _{zx}\mathbf{(\mathbf{r},\mathbf{r}^{\prime },}\omega \mathbf{)} &=&
\mathbf{\int }d\mathbf{\mathbf{k}_{\bot }\;}\frac{1}{(2\pi )^{2}}e^{i\mathbf{
k}_{\bot }(\mathbf{r}-\mathbf{r}^{\prime })_{\bot }}\left( -i\right) \frac{k
}{\epsilon \epsilon ^{\prime }}\frac{\partial }{\partial z^{\prime }}\left( 
\frac{e^{-\kappa \mid z-z^{\prime }\mid }+r^{\prime }e^{-\kappa (z+z^{\prime
})}}{2\kappa }\right) .  \label{Gzx}
\end{eqnarray}
Hence, for the first term in Eq.~(\ref{FFord1}) we have:

\begin{equation}
i\left\langle E^{j}\partial _{j^{\prime }}E^{z}\right\rangle =\partial
_{x^{\prime }}\Gamma ^{xz}+\partial _{y^{\prime }}\Gamma ^{yz}+\partial
_{z^{\prime }}\Gamma ^{zz}\,.  \label{Eproduct1}
\end{equation}
Next we drop the $\delta$-function terms, and take the coincidence limit,
$\mathbf{r}^\prime = \mathbf{r}$ and $z'=z$, after performing the
$\partial _{j^\prime }$ differentiation.
All derivatives in $y'$ are zero, since $\mathbf{k}_{\perp }$  
points along the $+x$ axis, so the second term above is zero. Using 
Eqs.~(\ref{Gxz}) and (\ref{Gzz}), we find the remaining terms 
(with $\epsilon =\epsilon ^{\prime }=1$ for the
vacuum region):

\begin{eqnarray}
\partial _{x^{\prime }}\Gamma ^{xz} &=&-\frac{1}{2}\mathbf{\int }d\mathbf{
\mathbf{k}_{\bot }\;}\frac{1}{(2\pi )^{2}}k^{2}\left( r^{\prime }e^{-2\kappa
z}\pm 1\right) ,  \label{Eproduct1term1} \\
\partial _{z^{\prime }}\Gamma ^{zz} &=&-\frac{1}{2}\mathbf{\int }d\mathbf{
\mathbf{k}_{\bot }\;}\frac{1}{(2\pi )^{2}}k^{2}\left( r^{\prime }e^{-2\kappa
z}\mp 1\right) \,.  \label{Eproduct1term2}
\end{eqnarray}
Here the sign of the last term is determined by whether $z$ approaches $z'$
from above (upper sign) or from below (lower sign). We will argue later that these
terms with ambiguous sign do not contribute to the final result.
Now Eq.~(\ref{Eproduct1}) becomes:

\begin{equation}
i\left\langle E^{j}\partial _{j^\prime}
E^{z}\right\rangle =-\mathbf{\int }d\mathbf{
\mathbf{k}_{\bot }\;}\frac{1}{(2\pi )^{2}}k^{2}r^{\prime }e^{-2\kappa z}.
\label{Eproduct1f}
\end{equation}
For the second term in Eq.~(\ref{FFord1}) we have:

\begin{equation}
i\left\langle E_{j}\partial ^{z^{\prime }}E^{j}\right\rangle =\partial
_{z^{\prime }}\left( \Gamma _{xx}+\Gamma _{yy}+\Gamma _{zz}\right) .
\label{Eproduct2}
\end{equation}
Using Eqs.~(\ref{Gxx}), (\ref{Gyy}), and (\ref{Gzz}), we can write:

\begin{eqnarray}
\partial _{z^{\prime }}\Gamma _{xx} &=&-\frac{1}{2}\mathbf{\int }d\mathbf{
\mathbf{k}_{\bot }\;}\frac{1}{(2\pi )^{2}}\kappa ^{2}\left( r^{\prime
}e^{-2\kappa z}\pm 1\right) ,  \label{Eproduct2term1} \\
\partial _{z^{\prime }}\Gamma _{yy} &=&-\frac{1}{2}\mathbf{\int }d\mathbf{
\mathbf{k}_{\bot }\;}\frac{1}{(2\pi )^{2}}\omega ^{2}\left( re^{-2\kappa
z}\mp 1\right) ,  \label{Eproduct2term2} \\
\partial _{z^{\prime }}\Gamma _{zz} &=&-\frac{1}{2}\mathbf{\int }d\mathbf{
\mathbf{k}_{\bot }\;}\frac{1}{(2\pi )^{2}}k^{2}\left( r^{\prime }e^{-2\kappa
z}\mp 1\right) ,  \label{Eproduct2term3}
\end{eqnarray}
so that Eq.~(\ref{Eproduct2}) becomes:

\begin{equation}
i\left\langle E_{j}\partial ^{z^\prime}E^{j}
\right\rangle =-\frac{1}{2}\mathbf{\int 
}d\mathbf{\mathbf{k}_{\bot }\;}\frac{1}{(2\pi )^{2}}\left[ \left(
2k^{2}-\omega ^{2}\right) r^{\prime }e^{-2\kappa z}+\omega ^{2}re^{-2\kappa
z}\mp 2\omega ^{2}\right] \, .  \label{Eproduct2f}
\end{equation}
For the third term in Eq.~(\ref{FFord1}), we can write:

\begin{equation}
\left\langle B_{j}\partial ^{j^\prime}B^{z}\right\rangle =\left\langle
B_{x}\partial _{x^\prime}B_{z}\right\rangle +\left\langle B_{y}\partial
_{y^\prime}B_{z}\right\rangle +\left\langle B_{z}\partial _{z^\prime}
B_{z}\right\rangle .  \label{Bproduct3}
\end{equation}
The nonzero terms, using Eqs.~(\ref{MfieldG}) and (\ref{Gyy}), become:

\begin{eqnarray}
i\left\langle B_{x}\partial _{x^\prime}B_{z}\right\rangle &=&-\frac{1}{2}
\mathbf{\int }d\mathbf{\mathbf{k}_{\bot }\;}\frac{1}{(2\pi )^{2}}k^{2}\left(
re^{-2\kappa z}\pm 1\right) ,  \label{Bproduct3term1} \\
i\left\langle B_{z}\partial _{z^\prime}B_{z}\right\rangle &=&-\frac{1}{2}
\mathbf{\int }d\mathbf{\mathbf{k}_{\bot }\;}\frac{1}{(2\pi )^{2}}k^{2}\left(
re^{-2\kappa z}\mp 1\right) ,  \label{Bproduct3term2}
\end{eqnarray}
So, Eq.~(\ref{Bproduct3}) becomes:

\begin{equation}
i\left\langle B_{j}\partial ^{j^\prime}B^{z}\right\rangle =-\frac{1}{2}
\mathbf{\int }d\mathbf{\mathbf{k}_{\bot }\;}\frac{1}{(2\pi )^{2}}k^{2}\
2re^{-2\kappa z}.  \label{Bproduct3f}
\end{equation}
To find the fourth term in Eq.~(\ref{FFord1}), note that \cite{Vasilka}:

\begin{eqnarray}
& & i\left\langle B_{j}B_{j}\right\rangle _{\omega }=\int \frac{d\mathbf{k}
_{\perp }}{(2\pi )^{2}}      \nonumber \\
&\times & \left[ \left( k^{2} + \nabla _{z}\nabla _{z^{\prime
}}\right) \left( \frac{e^{-\kappa \mid z-z^{\prime }\mid }+re^{-\kappa
(z+z^{\prime })}}{2\kappa }\right)     
+ \omega ^{2}\left( \frac{e^{-\kappa \mid
z-z^{\prime }\mid }+r^{\prime }e^{-\kappa (z+z^{\prime })}}{2\kappa }\right) 
\right] .  \label{Bproduct4}
\end{eqnarray}
From here we get:

\begin{equation}
i\left\langle B_{j}\partial ^{z^\prime}B^{j}\right\rangle =
i\partial _{z^\prime}
\left\langle B_{j}B_{j}\right\rangle =-\frac{1}{2}\mathbf{\int }d\mathbf{
\mathbf{k}_{\bot }\;}\frac{1}{(2\pi )^{2}}\left\{ \left[ \left(
2k^{2}-\omega ^{2}\right) r+\omega ^{2}r^{\prime }\right] e^{-2\kappa z}\mp
2\omega ^{2}\right\} .  \label{Bproduct4f}
\end{equation}
Combining Eqs.~(\ref{Eproduct1f}), (\ref{Eproduct2f}), (\ref{Bproduct3f}), 
and (\ref{Bproduct4f}), as in Eq.~(\ref{FFord1}), we have:

\begin{eqnarray}
& & 2\left\langle E^{j}\partial _{j^\prime}E^{z}\right\rangle +\left\langle
E_{j}\partial ^{z^\prime}E^{j}\right\rangle +
2\left\langle B_{j}\partial ^{j^\prime}B^{z}\right\rangle
 -2\left\langle B_{j}\partial ^{z^\prime}B^{j}\right\rangle 
 =                                        \nonumber \\
& & {\rm Re}\left(\frac{i}{2}\mathbf{\int }d\mathbf{\mathbf{k}_{\bot }\;}\frac{1}{
(2\pi )^{2}}\left\{ 3\left[ \left( 2k^{2}-\omega ^{2}\right) r^{\prime
}+\omega ^{2}r\right] e^{-2\kappa z}\pm 2\omega ^{2}\right\} \right)\, .
\label{EBproducts}
\end{eqnarray}
The last term above would give an infinite force and cannot be present. If
we average over $z>z^{\prime }$ and $z<z^{\prime }$, it would average to
zero. In the limit that $r = r' =0$, the force must vanish, so we can drop
the last term.
 Then, we let $\alpha \longrightarrow \alpha _{r}$, so that
 Eq.~(\ref{FFord1}) becomes:

\begin{equation}
F^{z}(\omega )={\rm Re}\left\{
\frac{i}{2}\mathbf{\int }d\mathbf{\mathbf{k}_{\bot }\;}\frac{
1}{(2\pi )^{2}}\alpha _{r}(\omega )\left[ \left( 2k^{2}-\omega ^{2}\right)
r^{\prime }+\omega ^{2}r\right] e^{-2\kappa z} \right\}
\end{equation}
This is equivalent to Eq.~(\ref{Mode}), with $\left\langle E^{2}\right\rangle
_{\omega }$ as defined in Eq.~(\ref{Efield}).

\end{document}